\documentclass[prl,aps,twocolumn,floatfix,showpacs]{revtex4}
\usepackage{graphics}
\begin{document}
\title{Two-species fermion mixtures with population imbalance}
\author{M. Iskin and C. A. R. S{\'a} de Melo}
\affiliation{School of Physics, Georgia Institute of Technology, Atlanta, Georgia 30332, USA}
\date{\today}

\begin{abstract}
We analyze the phase diagram of uniform superfluidity for
two-species fermion mixtures from the Bardeen-Cooper-Schrieffer (BCS) to 
Bose-Einstein condensation (BEC) limit as a function of the scattering parameter and
population imbalance. We find at zero temperature that the phase diagram of population imbalance 
versus scattering parameter is asymmetric for unequal masses, 
having a larger stability region for uniform superfluidity 
when the lighter fermions are in excess. 
In addition, we find topological quantum phase 
transitions associated with the disappearance or appearance of momentum space 
regions of zero quasiparticle energies. 
Lastly, near the critical temperature, we derive the Ginzburg-Landau 
equation, and show that it describes a dilute mixture of composite bosons 
and unpaired fermions in the BEC limit. 

\end{abstract}
\maketitle

Major experimental breakthroughs have been made recently
involving one-species trapped fermions ($^6$Li) in two hyperfine states
with different populations. The superfluid to normal 
phase transition and the vortex state~\cite{mit}, as well as phase separation between paired and 
unpaired fermions~\cite{rice} were identified as a function of population imbalance
and scattering parameter. These studies are important extensions 
of the so-called Bardeen-Cooper-Schrieffer (BCS) to Bose-Einstein condensation (BEC) 
evolution for equal populations, which were studied via the use of 
Feshbach resonances~\cite{coldatom1,coldatom2,coldatom3,coldatom4,coldatom5,coldatom6}. 
The problem of fermion superfluidity with population imbalance has
been revisited recently in several theoretical works in 
continuum~\cite{liu,bedaque,carlson,pao,son,sheehy,ho,bulgac,cheng,hui,sachdev} and
trapped~\cite{machida,pieri,torma,yi,chevy,silva,haque} atoms. 
These one-species experiments are ideal candidates for the observation of 
uniform and non-uniform superfluid phases, which may be present 
not only in atomic, but also in nuclear (pairing in nuclei), 
astrophysics (neutron stars), and condensed matter (superconductors) systems.

Arguably one of the next frontiers of exploration in ultracold Fermi 
systems is the search for superfluidity in two-species fermion mixtures 
(e.g. $^6$Li and $^{40}$K) with and without population imbalance.
While earlier works on two-species fermion mixtures were limited to the
BCS limit~\cite{liu,bedaque}, in this manuscript we study the evolution of superfluidity from the BCS to the 
BEC limit as a function of the scattering parameter and population imbalance.

\begin{figure} [htb]
\centerline{\scalebox{0.33}{\includegraphics{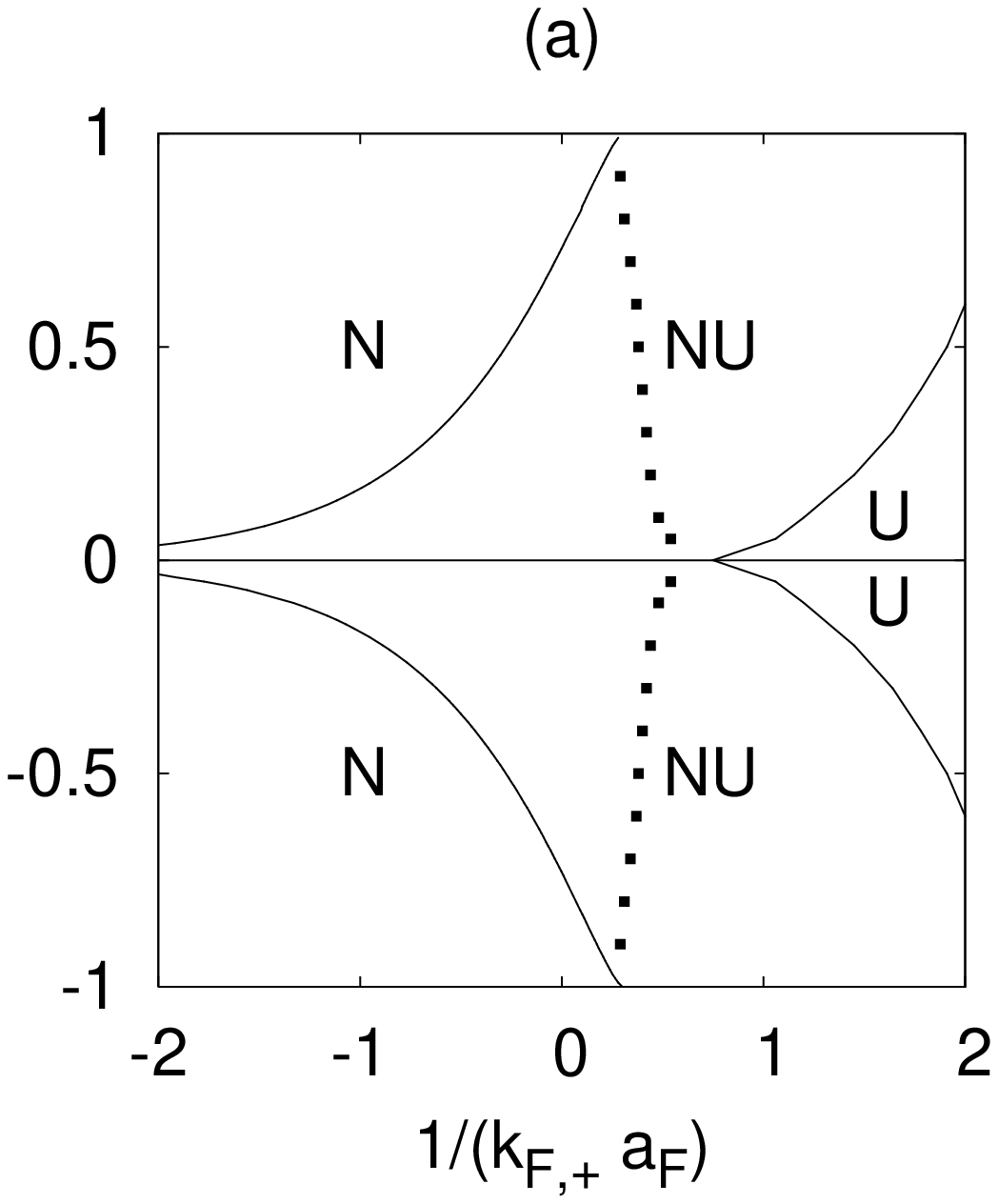} \includegraphics{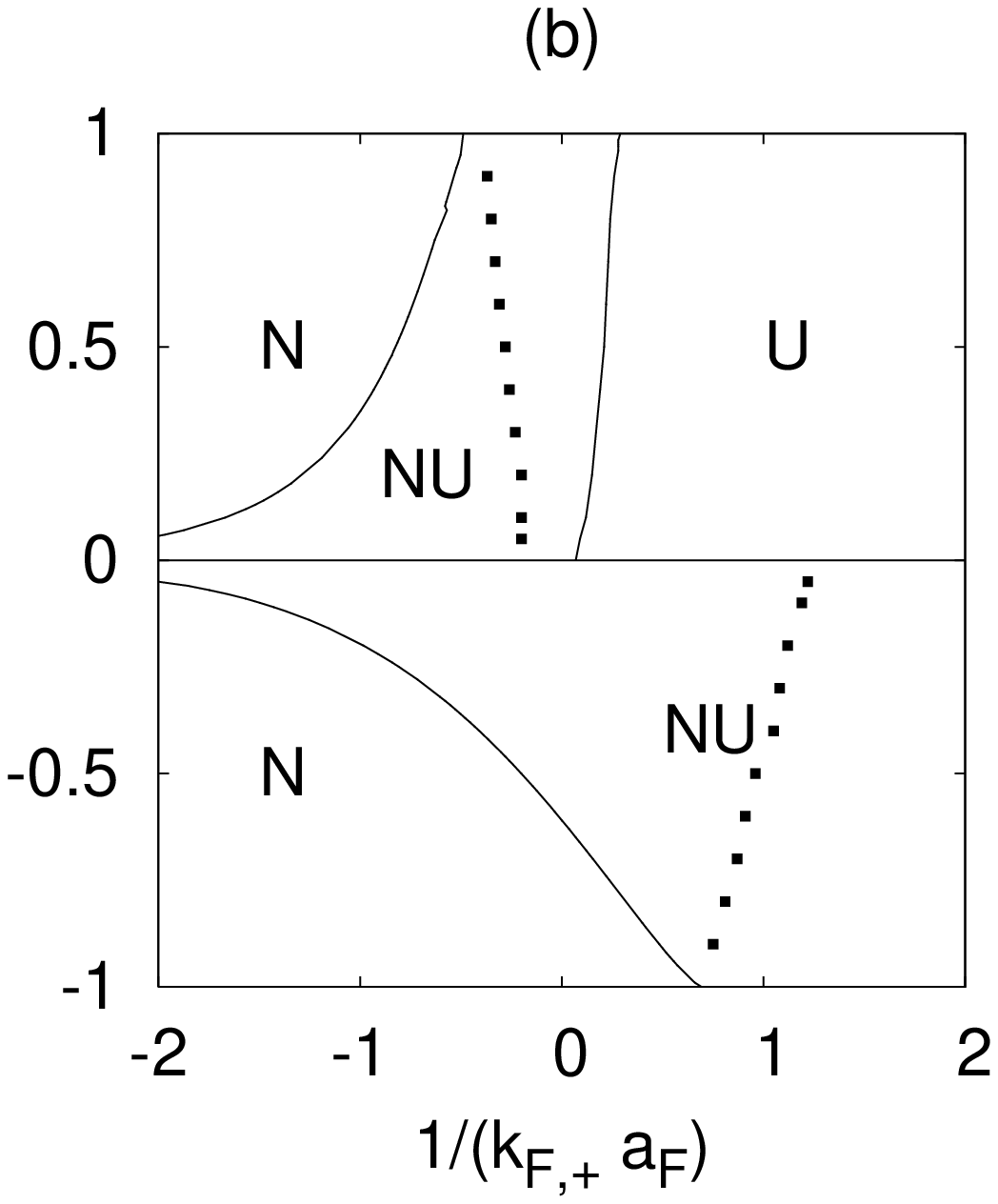}}}
\caption{\label{fig:phase} Phase diagram of 
$P = (N_\uparrow - N_\downarrow) / (N_\uparrow + N_\downarrow)$ versus $1/(k_{F,+} a_F)$ for
a) equal ($m_\uparrow = m_\downarrow$) and 
b) unequal masses ($m_\uparrow = 0.15 m_\downarrow$).
We show normal (N), non-uniform (NU) or uniform (U) superfluid phases.
The dotted and $P = 0$ lines separate topologically distinct regions. In b) the U phase
also occurs for $P <0$ when $1/(k_{F,+} a_F) > 4.8$ (not shown).
}
\end{figure}

Our main results are as follows. 
At zero temperature, we construct the phase diagram
for equal and unequal masses of paired fermions as a function of 
scattering parameter and population imbalance $P = (N_\uparrow - N_\downarrow) / (N_\uparrow + N_\downarrow)$ 
as shown in Fig.~\ref{fig:phase}. The phase diagram 
is asymmetric for unequal masses, having a larger 
stability region for uniform superfluidity when the population $N_\uparrow$ of lighter fermions  
is larger than the population $N_\downarrow$ of heavier fermions. 
In addition, we find topological quantum phase transitions in the phase
diagram associated with the disappearance or appearance of momentum space regions 
of zero quasiparticle energies. 
Lastly, near the critical temperature, we derive the time-dependent Ginzburg-Landau (TDGL)
equation, and show that it describes a dilute mixture of bosons (tightly bound fermions) 
and excess (unpaired) fermions in the BEC limit. 

To describe a dilute two-species Fermi gas in three dimensions, 
we start from the Hamiltonian ($\hbar = 1$)
\begin{equation}
\label{eqn:hamiltonian}
H = \sum_{\mathbf{k},\sigma} \xi_{\mathbf{k},\sigma} a_{\mathbf{k},\sigma}^\dagger a_{\mathbf{k},\sigma} + 
\sum_{\mathbf{k},\mathbf{k'},\mathbf{q}} V (\mathbf{k},\mathbf{k'})
b_{\mathbf{k},\mathbf{q}}^\dagger b_{\mathbf{k'},\mathbf{q}}, 
\end{equation}
where the pseudo-spin $\sigma$ labels the hyperfine states
represented by the creation operator $ a_{\mathbf{k},\sigma}^\dagger$, and
$b_{\mathbf{k},\mathbf{q}}^\dagger = a_{\mathbf{k}+\mathbf{q}/2,\uparrow}^\dagger 
a_{-\mathbf{k}+\mathbf{q}/2,\downarrow}^\dagger$.
Here, $\xi_{\mathbf{k},\sigma}= \epsilon_{\mathbf{k},\sigma} - \mu_\sigma$,
where $\epsilon_{\mathbf{k},\sigma} = k^2/(2m_\sigma)$ is the energy
and $\mu_\sigma$ is the chemical potential of the fermions.
Notice that, we allow for the fermions to have different masses $m_{\sigma}$ and
different populations controlled by independent chemical potentials $\mu_{\sigma}$.
The attractive fermion-fermion interaction $V (\mathbf{k},\mathbf{k'})$ 
can be written in a separable form as
$
V (\mathbf{k},\mathbf{k'}) =  - g \Gamma^*_\mathbf{k} \Gamma_\mathbf{k'} 
$
where $g > 0$, and $\Gamma_\mathbf{k} = 1$ for the s-wave symmetry.

The gaussian effective action for $H$ is~\cite{popov}
$
S_{\rm gauss} = S_0 + (\beta/2) \sum_{q} \bar{\Lambda}^\dagger(q) \mathbf{F}^{-1}(q) \bar{\Lambda}(q),
$
where $q=(\mathbf{q},v_\ell)$ with bosonic Matsubara frequency $v_\ell=2\ell\pi/\beta$.
Here, $\beta = 1/T$, $\bar{\Lambda}^\dagger(q)$ is the order parameter fluctuation field, 
and the matrix $\mathbf{F}^{-1}(q)$ is the inverse fluctuation propagator. 
The saddle point action is
\begin{eqnarray}
S_0 = \beta \frac{|\Delta_0|^2}{g} &+& \sum_\mathbf{k}\big\lbrace
\beta(\xi_{\mathbf{k},+} - E_{\mathbf{k},+}) \nonumber \\
&+& 
\ln [n_F(-E_{\mathbf{k},\uparrow})] + \ln [n_F(-E_{\mathbf{k},\downarrow})] 
\big\rbrace, 
\end{eqnarray}
where
$
E_{\mathbf{k},\sigma} = (\xi_{\mathbf{k},+}^2 + |\Delta_\mathbf{k}|^2)^{1/2} + s_\sigma \xi_{\mathbf{k},-}
$
is the quasiparticle energy when $s_\uparrow = 1$ and the negative of the quasihole energy when
$s_\downarrow = -1$, and
$
E_{\mathbf{k},\pm} = (E_{\mathbf{k},\uparrow} \pm E_{\mathbf{k},\downarrow})/2.
$
Here,
$
\Delta_\mathbf{k} = \Delta_0\Gamma_\mathbf{k}
$
is the order parameter,
$
n_F(E_{\mathbf{k},\sigma})
$
is the Fermi distribution and
$
\xi_{\mathbf{k},\pm} = (\xi_{\mathbf{k},\uparrow} \pm \xi_{\mathbf{k},\downarrow})/2
= k^2/(2m_\pm) - \mu_\pm,
$
where 
$
m_\pm = 2m_\uparrow m_\downarrow/(m_\downarrow \pm m_\uparrow)
$
and 
$
\mu_\pm = (\mu_\uparrow \pm \mu_\downarrow)/2.
$
Notice that $m_+$ is twice the reduced mass of the $\uparrow$ 
and $\downarrow$ fermions, and that the equal mass case corresponds 
to $|m_-| \to \infty$. 
The fluctuation term in the action leads to a correction
to the thermodynamic potential, which can be written as 
$\Omega_{{\rm gauss}} = \Omega_0 + \Omega_{{\rm fluct}}$ with 
$\Omega_0 = S_0/\beta$ and
$\Omega_{{\rm fluct}} = (1/\beta)\sum_{q}\ln\det[\mathbf{F}^{-1}(q)/\beta]$.

\begin{figure} [htb]
\centerline{\scalebox{0.33}{\includegraphics{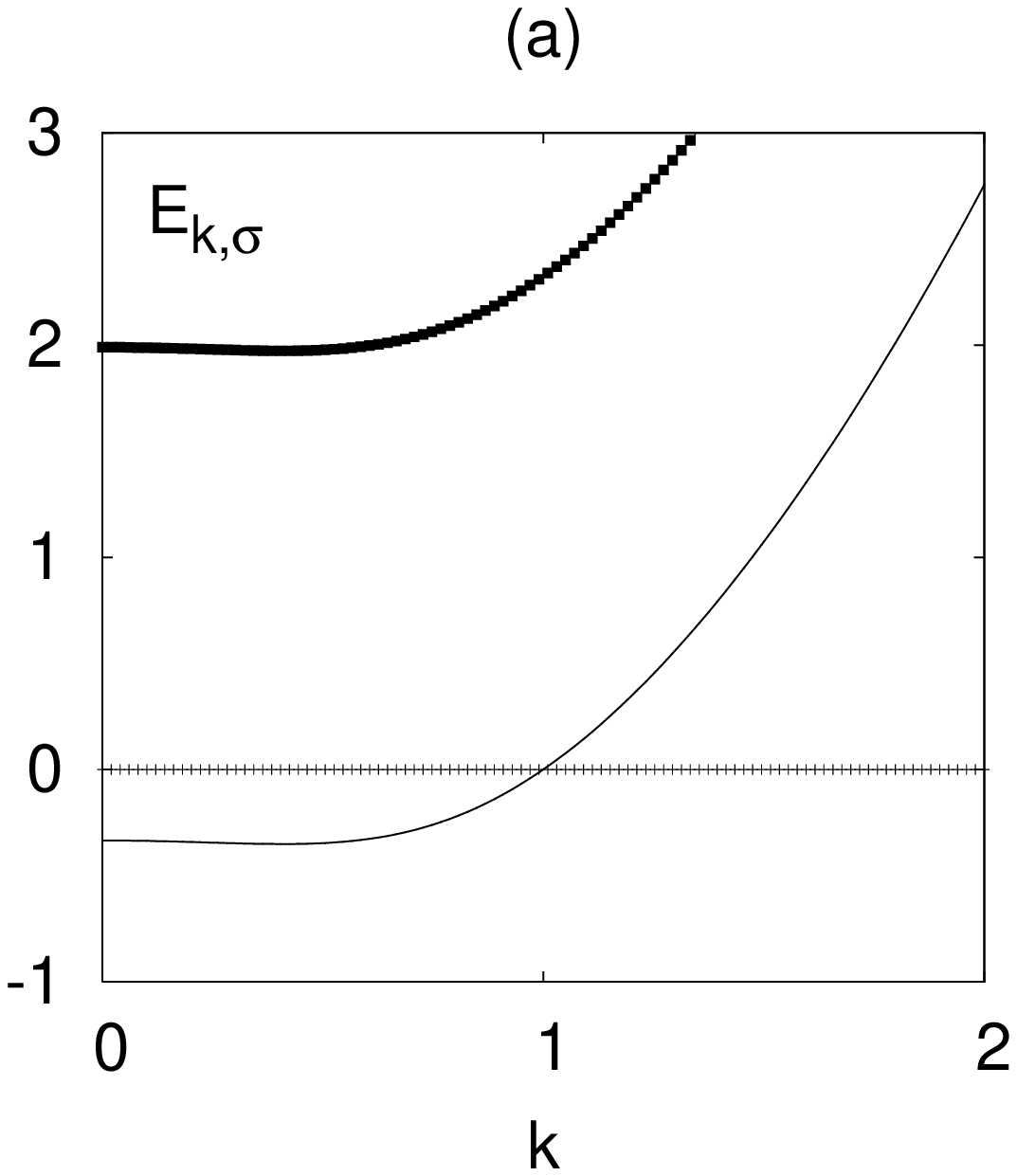} \includegraphics{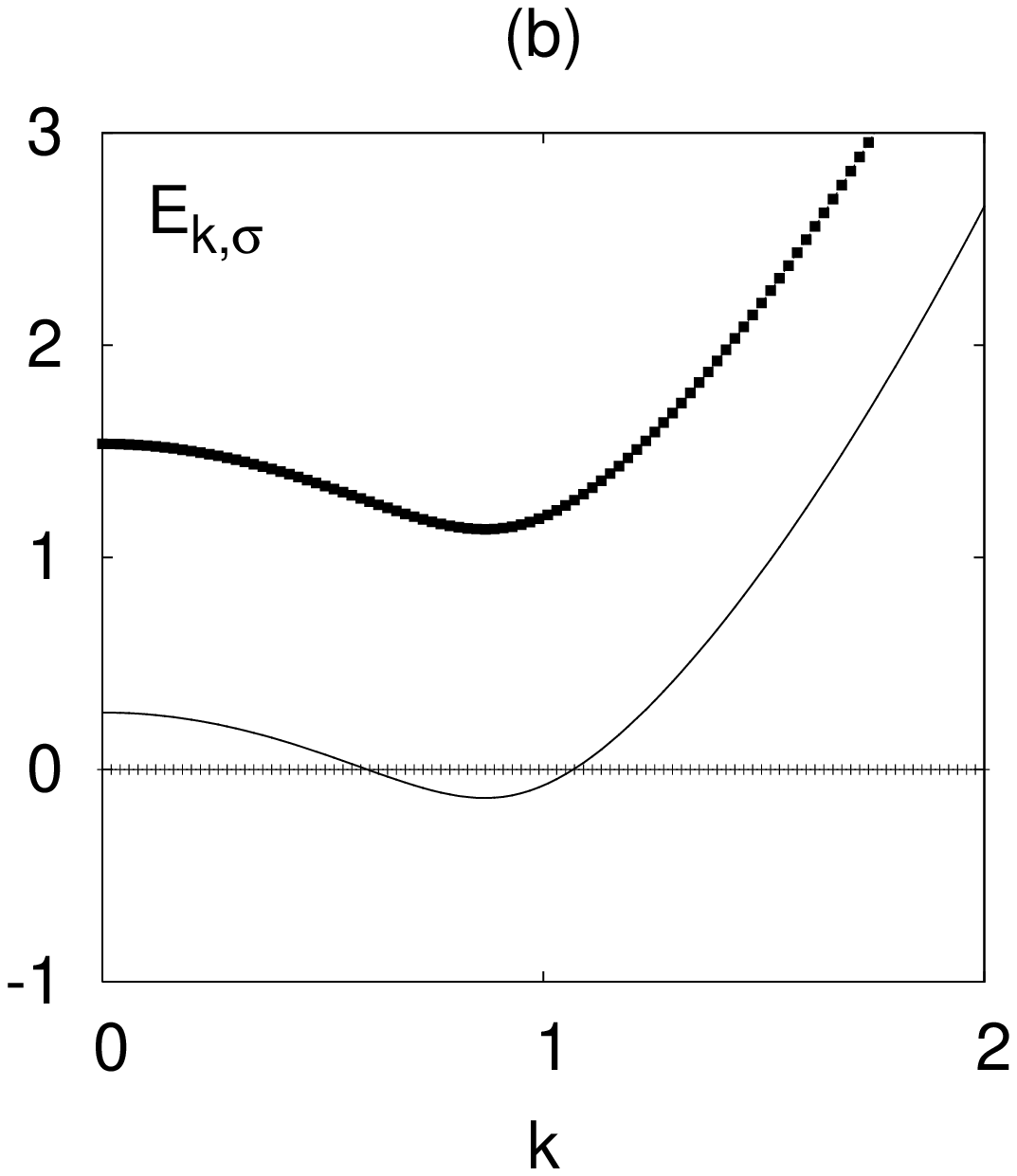}}}
\caption{\label{fig:md}
Schematic plots of $E_{\mathbf{k},\uparrow}$ (dotted lines)
and $E_{\mathbf{k},\downarrow}$ (solid lines) versus $k$
a) for case (I), and
b) for case (II).
}
\end{figure}

The saddle point condition $\delta S_0 /\delta \Delta_0^* = 0$ leads 
to an equation for the order parameter
\begin{equation}
\frac{1}{g} = \sum_{\mathbf{k}} \frac{|\Gamma_\mathbf{k}|^2} {2E_{\mathbf{k},+}} {\cal X}_{\mathbf{k},+},
\label{eqn:opeqn}
\end{equation}
where
$
{\cal X}_{\mathbf{k},\pm} = ( {\cal X}_{\mathbf{k},\uparrow} \pm {\cal X}_{\mathbf{k},\downarrow} ) / 2
$
with
$
{\cal X}_{\mathbf{k},\sigma} = \tanh(\beta E_{\mathbf{k},\sigma}/2).
$
As usual, we eliminate $g$ in favor of the scattering length $a_F$ via the relation
$
1/g = - m_+ V /(4\pi a_F) + \sum_{\mathbf{k}} |\Gamma_\mathbf{k}|^2 / (2\epsilon_{\mathbf{k},+}),
$
where 
$
\epsilon_{\mathbf{k},\pm} = (\epsilon_{\mathbf{k},\uparrow} \pm \epsilon_{\mathbf{k},\downarrow})/2.
$
The order parameter equation has to be solved self-consistently with number equations
$N_\sigma = -\partial \Omega/\partial {\mu_\sigma}$ which have two contributions 
$N_\sigma = N_{0,\sigma} + N_{{\rm fluct},\sigma}$.
$N_{0,\sigma}$ is the saddle point number equation given by
\begin{equation}
N_{0,\sigma} = - \frac{\partial \Omega_0} {\partial {\mu_\sigma}} 
= \sum_{\mathbf{k}} 
\left( \frac{1 - s_\sigma {\cal X}_{\mathbf{k},-}} {2}
- \frac{\xi_{\mathbf{k},+}}{2E_{\mathbf{k},+}} {\cal X}_{\mathbf{k},+}
\right)
\label{eqn:numbereqn}
\end{equation}
and $N_{{\rm fluct},\sigma} = -\partial \Omega_{{\rm fluct}}/\partial {\mu_\sigma}$ is the 
fluctuation contribution to $N$ given by
$
N_{{\rm fluct},\sigma} = - (1/\beta) \sum_{q} \lbrace 
\partial [\det \mathbf{F}^{-1}(q)] / \partial {\mu_\sigma} 
\rbrace / \det \mathbf{F}^{-1}(q).
$
We define 
$
B_\pm = m_\uparrow \mu_\uparrow \pm m_\downarrow \mu_\downarrow
$
to establish general constraints on the magnitude $|\Delta_0|$ of the order parameter
for s-wave pairing in the presence of population imbalance ($N_\uparrow \ne N_\downarrow$).
Population imbalance is achieved when either $E_{\mathbf{k},\uparrow}$
or $E_{\mathbf{k},\downarrow}$ is negative in some regions of momentum space. 
Depending on the number of zeros of $E_{\mathbf{k},\uparrow}$ and $E_{\mathbf{k},\downarrow}$
(zero energy surfaces in momentum space), there are two topologically distinct cases:
(I) $E_{\mathbf{k},\sigma}$ has no zeros and $E_{\mathbf{k},-\sigma}$ has only one, and
(II) $E_{\mathbf{k},\sigma}$ has no zeros and $E_{\mathbf{k},-\sigma}$ has two zeros.
The zeros of $E_{\mathbf{k},\sigma}$ occur at real momenta
$
k_{\pm}^2 = B_+ \pm (B_-^2 - 4m_\uparrow m_\downarrow |\Delta_0|^2)^{1/2}
$
provided that 
$
|\Delta_0|^2 < |B_-|^2/(4m_\uparrow m_\downarrow)
$
for $B_+ \ge 0$ and 
$
|\Delta_0|^2 < - \mu_\uparrow \mu_\downarrow
$
for $B_+ < 0$.
The $P  = 0$ limit corresponds to case (III), 
where $E_{\mathbf{k},\sigma}$ has no zeros and is always positive.
We illustrate these cases in Fig.~\ref{fig:md} for $N_\uparrow > N_\downarrow$.
Notice that, the Fermi sea of the lower quasiparticle band is a sphere of radius $k_+$ in case (I), 
and is a spherical shell $k_- \le k \le k_+$ in case (II). 
The transition from case (II) to case (I) occurs when $k_- \to 0$,
indicating a change in topology in the lowest quasiparticle band, 
similar to the Lifshitz transition in ordinary metals and 
non-swave superfluids~\cite{volovik,duncan}. The topological transition
here is unique, because it involves an s-wave superfluid, and could be potentially
observed for the first time through the measurement of the momentum distribution.

\begin{figure} [htb]
\centerline{\scalebox{0.33}{\includegraphics{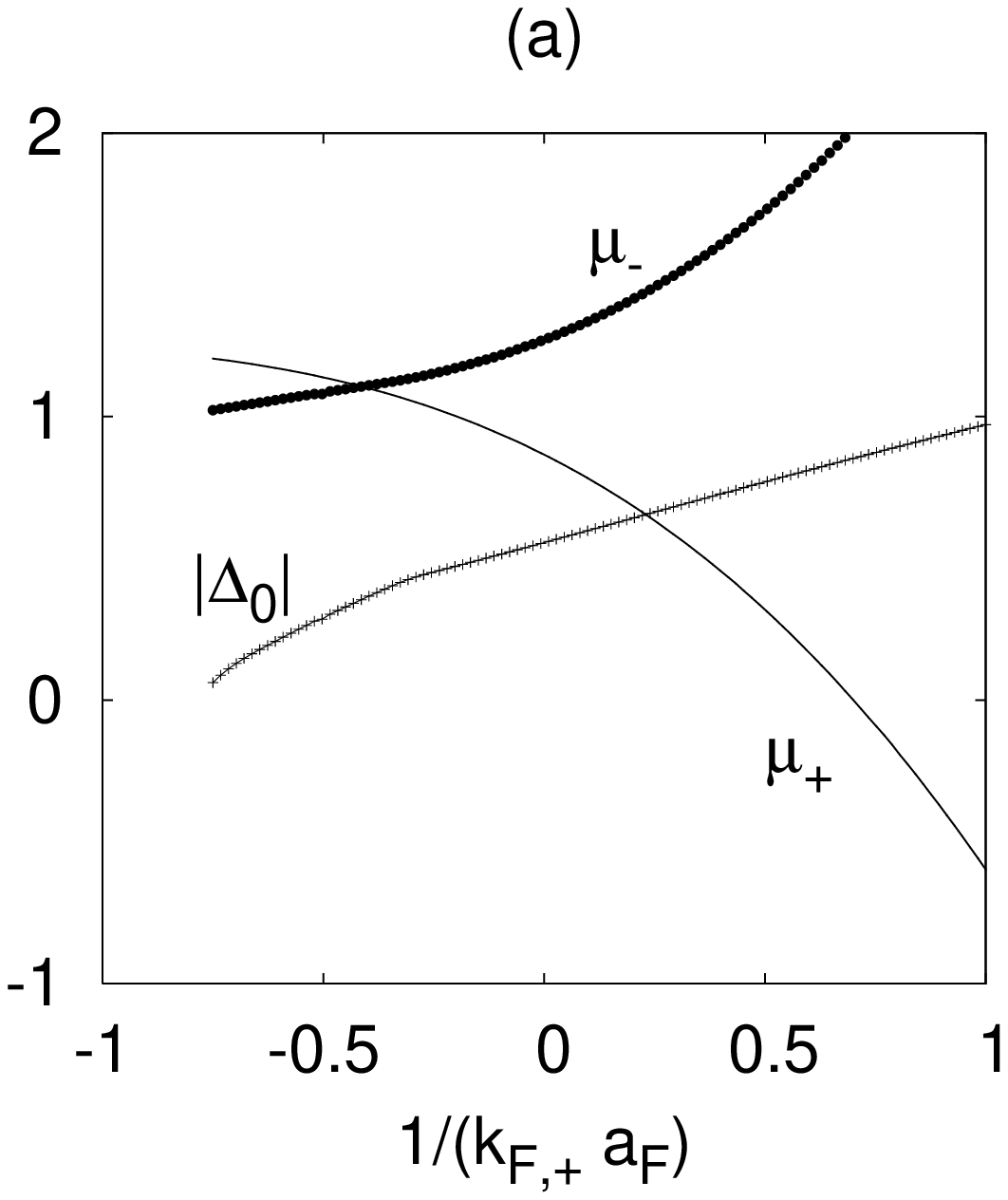} \includegraphics{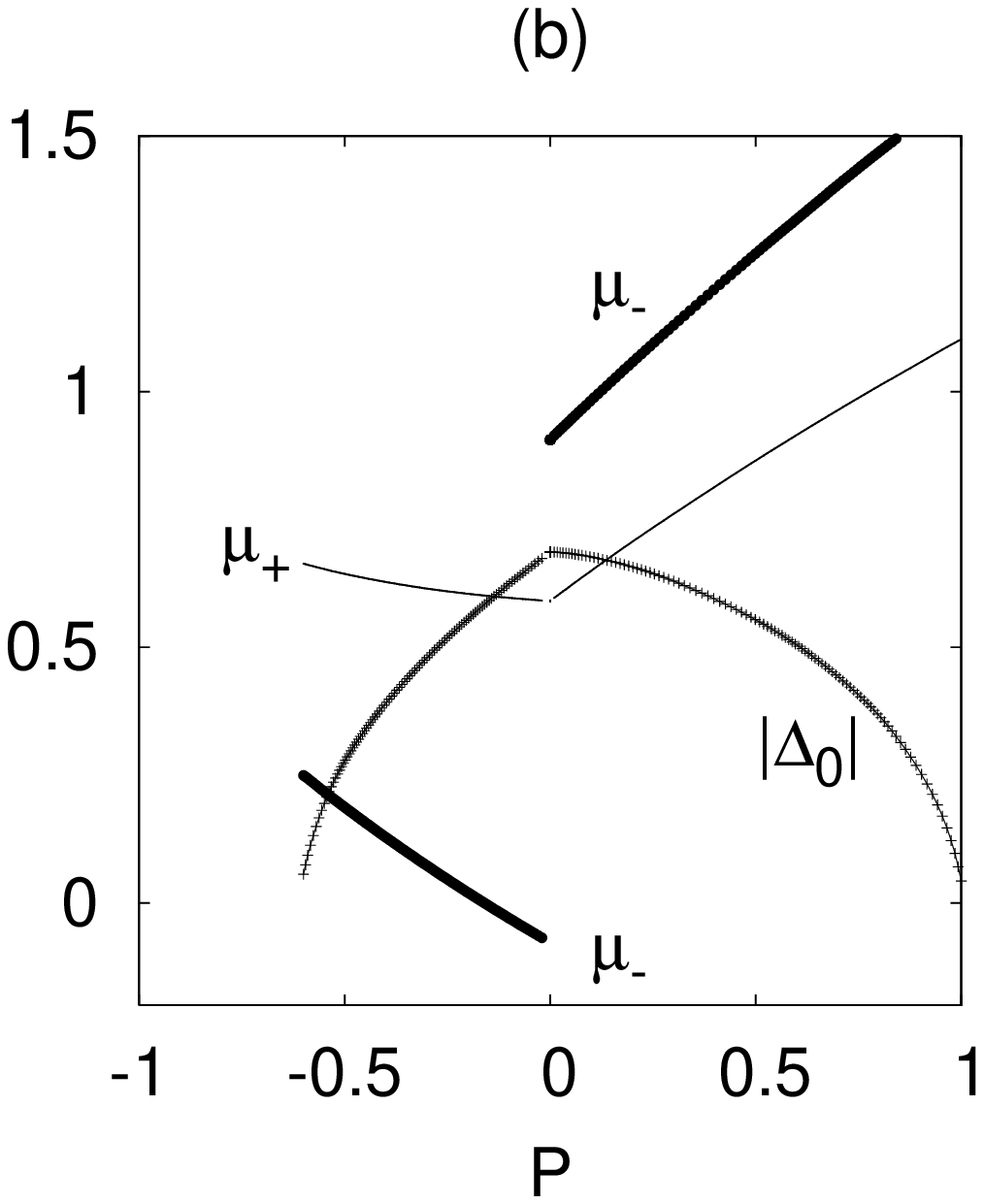}}}
\caption{\label{fig:gap} 
Plots of $|\Delta_0|$, $\mu_+$ and $\mu_-$ (units of $\epsilon_{F,+}$)
for $m_\uparrow = 0.15m_\downarrow$
a) as a function of $1/(k_{F,+} a_F)$ when $P = 0.5$ and
b) as a function of $P$ when $1/(k_{F,+} a_F) = 0$.
}
\end{figure}

The $T=0$ momentum distributions for cases (I) and (II) can be obtained from Eq.~\ref{eqn:numbereqn}.
For momentum space regions where $E_{{\bf k},\sigma} > 0$ and $E_{{\bf k},-\sigma} > 0$,
the corresponding momentum distributions are equal $n_{{\bf k},\sigma} =  n_{{\bf k},-\sigma}$.
However, when $E_{{\bf k},\sigma} > 0$ and $E_{{\bf k},-\sigma} < 0$, then
$n_{{\bf k},\sigma} = 0$ and $n_{{\bf k},-\sigma} = 1$. 
Although this topological transition
is quantum ($T=0$) in nature, signatures of the transition should still be observed
at finite temperatures within the quantum critical region, where the momentum distributions
are smeared out due to thermal effects. Although the primary signature of this topological 
transition is seen in the momentum distribution, the isentropic $\kappa_S$ or isothermal $\kappa_T$ compressibilities 
and the speed of sound $c_s$ would have a cusp at the topological transition line 
similar to that encountered in $\vert \Delta_0 \vert$ (see Fig.~\ref{fig:gap})
as a function of the scattering parameter $1 / (k_{F,+} a_F) $. The cusp (discontinuous change in
slope) in $\kappa_S$, $\kappa_T$ or $c_s$ gets larger with increasing population imbalance.

Next, we solve Eqs~(\ref{eqn:opeqn}) and~(\ref{eqn:numbereqn}) to 
analyze the phase diagram at $T = 0$ as a function of
scattering parameter $1/(k_{F,+} a_F)$ and population imbalance $P = N_-/N_+$,
where $N_\pm = (N_\uparrow \pm N_\downarrow)/2$ and 
$k_{F,\pm}^3 = (k_{F,\uparrow}^3 \pm k_{F,\downarrow}^3)/2$.
We perform calculations for equal ($m_\uparrow = m_\downarrow$) and unequal 
($m_\uparrow = 0.15 m_\downarrow$) masses cases, corresponding to one-species
($^6$Li or $^{40}$K only), and two-species ($^6$Li and $^{40}$K mixture) experiments, respectively.

For $T \approx 0$, $N_{{\rm fluct},\sigma}$ is small compared to 
$N_{0,\sigma}$ for all $1/(k_{F,+} a_F)$ leading to $N_\sigma \approx N_{0,\sigma}$~\cite{jan}.
We define $\epsilon_{\rm F,\pm} = k_{F,\pm}^2/(2m_{\pm})$ and
in  Fig.~\ref{fig:gap}, we plot self-consistent solutions of
$|\Delta_0|$, $\mu_+$ and $\mu_-$ (in units of $\epsilon_{F,+}$) at $T = 0$ for two cases:
a) as a function of $1/(k_{F,+} a_F)$ when $P = 0.5$ (or $N_\uparrow = 3N_\downarrow$) and
b) as a function of $P$ when $1/(k_{F,+} a_F) = 0$ (or on resonance).

In Fig.~\ref{fig:gap}a, the BCS $\mu_\pm \approx (\epsilon_{F,\uparrow} \pm \epsilon_{F,\downarrow})/2$ 
changes continuously to the BEC $|\mu_\pm| \to |\epsilon_b|/2$,
where $\epsilon_b = -1/(m_+ a_F^2)$ is the binding energy which follows from
$
1/g = \sum_{\mathbf{k}} |\Gamma_\mathbf{k}|^2/(2\epsilon_{\mathbf{k},+} - \epsilon_b).
$
Since $P > 0$ and the $\uparrow$ fermions are in excess, all $\downarrow$ fermions pair
to form $N_\downarrow$ bosons and the remaining $\uparrow$ fermions are unpaired.
The amplitude $|\Delta_{0}|$ evolves continuously from the BCS to the BEC limit
with a cusp around $1/(k_{F,+} a_F) \approx - 0.27$.
This cusp in $|\Delta_0|$ is more pronounced for higher $|P|$ and signals a quantum phase transition 
from case (I) to case (II), which can be detected experimentally~\cite{coldatom6}.

In Fig.~\ref{fig:gap}b, we show (on resonance) that $|\Delta_0| = 0$ (normal phase)
for $P < -0.61$, where $\mu_+ \approx 0.64 \epsilon_{F,+}$ and $\mu_- \approx 0.28 \epsilon_{F,+}$.
We notice that the evolution of $|\Delta_0|$, $\mu_+$ and $\mu_-$ 
as a function of $P$ is non-analytic when $|P| \to 0$, and
signals a quantum phase transition from case (II) with $P > 0$ 
to case (III) with $P = 0$ to case (I) with $P < 0$.
We obtain similar results when $m_\uparrow = m_\downarrow$, where
the plot is symmetric around $P = 0$.
Therefore, this quantum phase transition may be studied in 
current experiments involving only one-species of fermions~\cite{mit,rice}.

Next, we discuss the stability of uniform superfluidity using two criteria.
The first criterion requires that the curvature
$\partial^2 \Omega_0 / \partial \Delta_0^2$ of the saddle point free energy
$\Omega_0 = S_0/\beta$ with respect to $\Delta_0$ to be positive.
When $\partial^2 \Omega_0 /\partial \Delta_0^2$ is
negative, the uniform saddle point solution does not correspond to 
a minimum of $\Omega_0$, and a non-uniform superfluid phase is favored.
The second criterion requires the eigenvalues of the superfluid density
\begin{equation}
\rho_{ij} (T) = (m_\uparrow N_\uparrow + m_\downarrow N_\downarrow)\delta_{ij}
- \frac{\beta}{2} \sum_{\mathbf{k}} k_i k_j {\cal Y}_{\mathbf{k},+}
\end{equation}
to be positive. Here, $\delta_{ij}$ is the Kronecker delta,
and ${\cal Y}_{\mathbf{k},\pm} = ({\cal Y}_{\mathbf{k},\uparrow} \pm  {\cal Y}_{\mathbf{k},\downarrow})/2$,
with ${\cal Y}_{\mathbf{k},\sigma} = {\rm sech}^2 (\beta E_{ \mathbf {k}, \sigma}/2)$. 
When at least one of the eigenvalues of $\rho_{ij}(T)$ is negative, a spontaneously generated gradient
of the phase of the order parameter appears, leading to a non-uniform superfluid phase.
Notice that, $\rho_{ij}(T)$ reduces to a scalar $\rho_0(T)$ for s-wave systems.

The uniform superfluid (U) phase is characterized by $\rho_0 (0) > 0$ and 
$\partial^2 \Omega_0 /\partial \Delta_0^2 > 0$, and the normal (N) phase 
is characterized by $\Delta_0 = 0$.
The non-uniform superfluid (NU) phase is characterized by
$\rho_0 (0) < 0$ and/or $\partial^2 \Omega_0 /\partial \Delta_0^2 < 0$. 
The NU phase should be of the LOFF-type having one wavevector modulation for the center
of mass momentum of a Cooper pair near the BCS limit,
where pairing occurs within a very narrow region around the lowest 
Fermi energy of either the lighter or heavier atom.  
Towards unitarity and beyond, we expect the NU phase to be substantially 
different from LOFF phases having spatial modulation that would encompass several
wavevectors. This expectation is based on the idea that when the attraction 
between fermions gets stronger, the Fermi mixture becomes more non-degenerate, and
there is a wider region in energy within which pairing can occur, leading to a 
range of possible wavevectors for the center of mass momentum of Cooper pairs. 

As shown in Fig.~\ref{fig:phase}a, when $m_\uparrow = m_\downarrow$, 
the phase diagram is symmetric around $P = 0$. 
A continuous quantum phase transition occurs 
from the NU to the N phase beyond a critical 
population imbalance on the BCS side. In addition, a discontinous transition 
from the NU to the U phase of topological type (I) also occurs.

In contrast, as shown in Fig.~\ref{fig:phase}b, when $m_\uparrow = 0.15 m_\downarrow$, 
the phase diagram is asymmetric around $P = 0$. 
A continuous quantum phase transition occurs from the NU 
to the N phase beyond a critical population imbalance
on the BCS side. Furthermore, it is found that the U phase has a larger stability
region when light fermions are in excess, and that a discontinuous transition from 
the NU to the U phase occurs. The U phase also exists for $P < 0$ when $1/(k_{F,+} a_F) > 4.8$ (not shown).
Lastly, one of the topological quantum phase transitions (dotted lines)
is very close to the NU/U boundary for $P >0$ in contrast to the equal
mass case. This line indicates a change in quasiparticle Fermi surface topology 
from type (I) to type (II), and may lie in the U region when
$m_{\uparrow}/m_{\downarrow} < 0.15$.

Next, we discuss superfluidity near $T \approx T_c$, 
where $\Delta_0 = 0$, and derive the TDGL equation~\cite{carlos}.
We use the small $\mathbf{q}$ and $iv_{\ell} \to \omega + i\delta$ expansion of 
$$
L^{-1}(q) = \frac{1}{g} - \sum_{\mathbf{k}} 
\frac{1 - n_F(\xi_{\mathbf{k} + \mathbf{q}/2,\uparrow}) - n_F(\xi_{\mathbf{k} - \mathbf{q}/2,\downarrow})} 
{\xi_{\mathbf{k} + \mathbf{q}/2,\uparrow} + \xi_{ \mathbf{k} - \mathbf{q}/2,\downarrow} - iv_\ell} |\Gamma_\mathbf{k}|^2,
$$
where $L^{-1}(q) = \mathbf{F}^{-1}_{11}(q)$, to obtain the
TDGL equation
\begin{equation}
\left[ a + b|\Lambda(x)|^2 - \sum_{i,j}\frac{c_{ij}}{2}\nabla_i\nabla_j - 
id\frac{\partial}{\partial t} \right]\Lambda(x) = 0
\end{equation}
in the real space $x = (\mathbf{x},t)$ representation.
The coefficients are given by
$
a = 1/g - \sum_{\mathbf{k}} X_{\mathbf{k},+}
|\Gamma(\mathbf{k})|^2 / (2\xi_{\mathbf{k},+}),
$
which leads to the saddle equation when $a = 0$ (Thouless condition), 
and
$
c_{ij} = \sum_{\mathbf{k}} \big\lbrace
\beta^2 k_i k_j 
	( X_{\mathbf{k},\uparrow} Y_{\mathbf{k},\uparrow}/m_\uparrow^2 
	+ X_{\mathbf{k},\downarrow} Y_{\mathbf{k},\downarrow}/m_\downarrow^2 ) / (32\xi_{\mathbf{k},+})
+ \beta (k_i k_j	C_- / (m_-\xi_{\mathbf{k},+}) - \delta_{ij}	C_+ / 2) (8 \xi_{\mathbf{k},+})
+ X_{\mathbf{k},+}
	[\delta_{ij}/(2m_+) - k_i k_j / (m_-^2\xi_{\mathbf{k},+})] / (4\xi_{\mathbf{k},+}^2)
\big\rbrace |\Gamma_\mathbf{k}|^2,
$
where 
$
C_\pm = (Y_{\mathbf{k},\uparrow}/m_\uparrow \pm Y_{\mathbf{k},\downarrow}/m_\downarrow) / 2,
$
$
X_{\mathbf{k},\sigma} = \tanh(\beta\xi_{\mathbf{k},\sigma}/2)
$ 
and 
$
Y_{\mathbf{k},\sigma} = {\rm sech}^2(\beta\xi_{\mathbf{k},\sigma}/2).
$
Notice that $c_{ij}$ reduces to a scalar $c$ in the s-wave case.
The coefficient of the nonlinear term is
$
b = \sum_{\mathbf{k}}[
X_{\mathbf{k},+} /(4\xi_{\mathbf{k},+}^3) 
- \beta Y_{\mathbf{k},+} / (8\xi_{\mathbf{k},+}^2) 
] |\Gamma_\mathbf{k}|^4,
$
while $d$ has real and imaginary parts given by
$
d = \lim_{w \to 0} \sum_{\mathbf{k}}
X_{\mathbf{k},+} 
[ 1/(8 \xi_{\mathbf{k},+}^2) + i \pi \delta(2\xi_{\mathbf{k},+} - w)/(2w)
] |\Gamma_\mathbf{k}|^2,
$
where $\delta(x)$ is the Delta function.
Notice that the damping term (imaginary part of $d$) vanishes 
for $\mu_+  \le 0$, indicating an undamped dynamics for $\Lambda(x)$.

Since a uniform superfluid phase is more stable in the BEC side, 
we calculate analytically all coefficients in the BEC limit
where $|\mu_\pm| \sim |\epsilon_b|/2 \gg T_{\rm c}$.
We obtain
$a = a_1 + a_2 = - V m_+^2 (2\mu_+ - \epsilon_b) a_F/(8\pi) + V m_+ n_e a_F^2$,
$b = b_1 + b_2 = V m_+^3 a_F^3/(16 \pi) - V m_+^2 (\partial n_e/\partial \mu_e) a_F^4$,
$c = V m_+^2 a_F / [8\pi(m_\uparrow + m_\downarrow)]$, and
$d = V m_+^2 a_F / (8\pi)$.
Here, $e$ labels the excess type of fermions and 
$n_e$ is the density of unpaired fermions.
Through the rescaling $\Psi(x) = \sqrt{d}\Lambda(x)$,
we obtain the equation of motion 
for a dilute mixture of weakly interacting bosons and fermions
\begin{eqnarray}
\mu_B \Psi(x) &+& \left[U_{BB}|\Psi(x)|^2 + U_{BF} n_e(x) \right] \Psi(x) \nonumber \\ 
&-& \frac{\nabla^2 \Psi(x)}{2m_B}  - i\frac{\partial \Psi(x)}{\partial t} = 0,
\label{eqn:GL-BEC}
\end{eqnarray}
with bosonic chemical potential 
$\mu_B = - a_1/d = 2\mu_+ - \epsilon_b$, 
mass $m_B = d/c = m_\uparrow + m_\downarrow$,
and repulsive boson-boson 
$U_{BB} = b_1 V/d^2 = 4\pi a_F / m_+$
and boson-fermion
$U_{BF} = a_1 V/(d n_e) = 8\pi a_F / m_+$ interactions.
This procedure also yields the spatial density of unpaired fermions given by
$
n_e(x) = [a_2/d + b_2|\Psi(x)|^2/d^2]/U_{BF} =
n_e - 8 \pi a_F (\partial n_e/\partial \mu_e) |\Psi(x)|^2/ m_+.
$
Since $\partial n_e/\partial \mu_e > 0$ the unpaired fermions avoid regions
where the boson field $|\Psi (x)|$ is large. Thus, the bosons condense
at the center and the unpaired fermions tend to be at the edges in a trap.
Notice that, Eq.~(\ref{eqn:GL-BEC}) reduces to the Gross-Pitaevskii equation 
for equal masses with $P = 0$~\cite{carlos},
and to the equation of motion for equal masses with $P \ne 0$~\cite{pieri}.

Furthermore, we obtain the boson-boson
$
a_{BB} = [1 + m_\uparrow/(2m_\downarrow) + m_\downarrow/(2m_\uparrow)] a_F
$
and boson-fermion
$
a_{BF} = 4 m_B m_e/[m_+ (m_B + m_e)] a_F
$
scattering lengths, which reduce to 
$a_{BB} = 2a_F$ and $a_{BF} = 8a_F/3$ for equal masses~\cite{carlos,pieri}.
For a mixture of $^6$Li and $^{40}$K, 
$a_{BB} \approx 4.41 a_F$, and
$a_{BF} \approx 2.03 a_F$ when $^6$Li is in excess, and 
$a_{BF} \approx 8.20 a_F$ when $^{40}$K is in excess.
For a better estimate, higher order scattering processes are needed~\cite{gora}. 
Since the effective boson-fermion system is weakly interacting, the BEC temperature is
$
T_c = \pi [n_B/\zeta(3/2)]^{2/3}/m_B,
$
where $\zeta(x)$ is the Zeta function and $n_B = (n - n_e)/2$.

In summary, we analyzed the phase diagram of uniform superfluidity for
two-species fermion mixtures (e.g. $^6$Li and $^{40}$K) from the BCS to 
the BEC limit as a function of scattering parameter and population imbalance.
We found that the zero temperature phase diagram of population imbalance versus scattering parameter 
is asymmetric for unequal masses, having a larger stability region for uniform superfluidity 
when the lighter fermions are in excess. This result is in sharp 
contrast with the symmetric phase diagram for equal masses. 
In addition, we found topological quantum phase transitions
associated with the disappearance or appearance of momentum space regions of zero quasiparticle 
energies. Near the critical temperature, we derived the Ginzburg-Landau 
equation, and showed that it describes a dilute mixture of bosons 
(tightly bound fermions) and excess (unpaired) fermions in the BEC limit. 

We thank NSF (DMR-0304380) for the support.

\end{document}